\documentclass[preprint]{aastex}

\newcommand{\gn}{g_{_{\rm N}}}
\newcommand{\Mn}{M_{_{\rm N}}}
\newcommand{\taun}{\tau_{_{\rm N}}}

\begin{document}

\title{MOND and the fundamental plane of elliptical galaxies}

\author{Riccardo Scarpa}
\affil{European Southern Observatory, 3107 Alonso de Cordova, 
Santiago, Chile} 
\email{rscarpa@eso.org}

\begin{abstract} 
It is shown that the MOdified Newtonian Dynamics (MOND)
explains the tilt of the fundamental plane of elliptical galaxies
without the need of non-baryonic dark matter. Results 
found for elliptical galaxies extends to
globular clusters and galaxy clusters, showing that MOND
agrees with observations over 7 order of magnitude in acceleration.
\end{abstract}

\section{Introduction}
\begin{figure}
\plotone{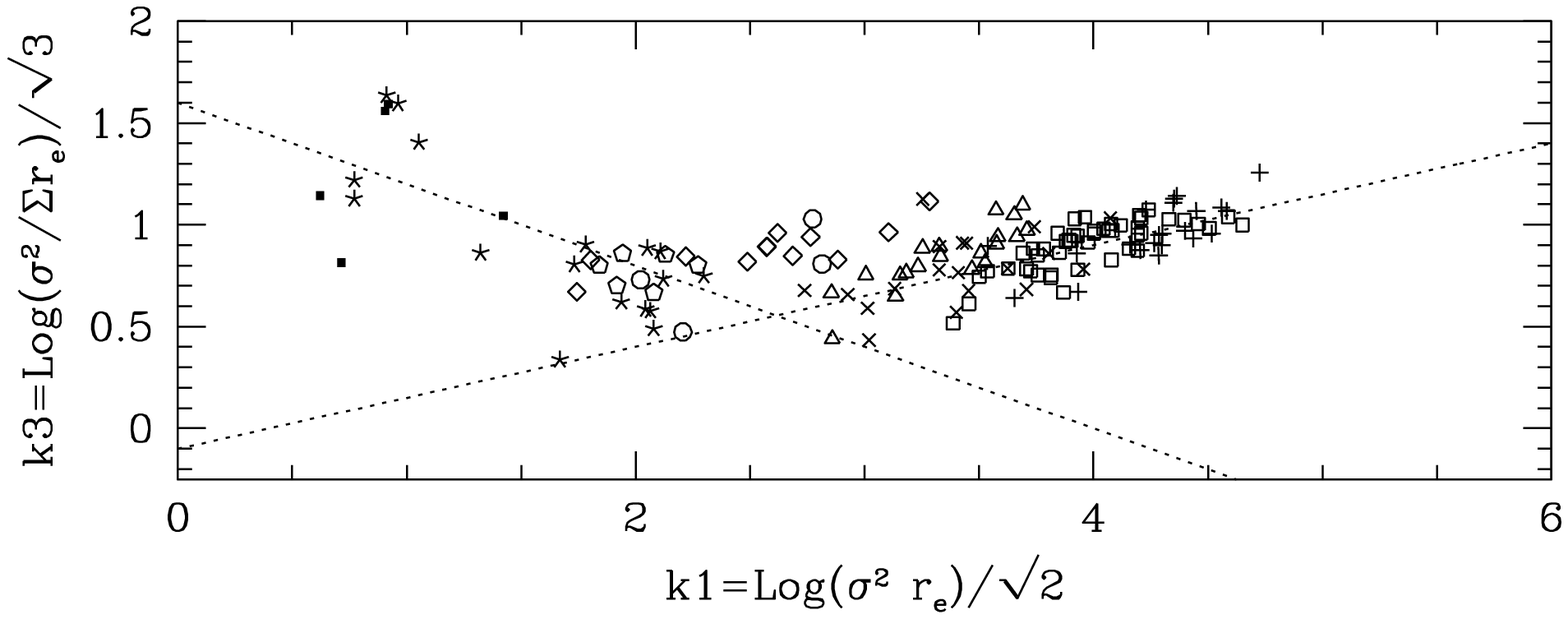}
\caption{Edge on view of the fundamental plane of elliptical 
galaxies. The quantities $k1$ and $k3$,
as defined by \citet{bender93}, are used. In Newtonian dynamics,
$k_1 \propto M$, while $k_3 \propto \tau$.
The two dotted lines  have slope $-0.4$ and $+0.25$, 
showing the dependence of $\tau$ on $M$ for LSB and HSB, respectively. 
Data are from \citet{bender93}: Squares= giant ellipticals; 
Triangles= intermediate ellipticals; Diamonds= dwarf ellipticals;
circle= compact ellipticals; Crosses=bulges; Dots= LSB ellipticals;
Stars= LSB from \citet{peterson93}; Pentagons= dwarf elliptical in Virgo
from \citet{marel02}; Pluses= radio galaxies from \citet{bettoni01}.
\label{FP}
}
\end{figure}

High-surface-brightness (HSB) elliptical galaxies are known to define
a ``fundamental plane'' (FP) when represented in the logarithmic space
of their effective radius $r_e$, average surface brightness $\Sigma$
within $r_e$, and central stellar velocity dispersion $\sigma$
(\cite{dressler87}; \cite{djorgovski87}; Fig. \ref{FP}).  If
elliptical galaxies had a constant mass-to-light ratio (M/L), then the
physical base for the FP would be obvious because the virial theorem
would imply a link between these three quantities.  Real galaxies,
however, do not follow the plane defined by the virial theorem
(\cite{busarello97}; \cite{levine96}; \cite{prugniel97}). Finding an
explanation for this unexpected behavior has been and still is one of
the major puzzle in modern astrophysics. In a nutshell, both
possibilities of a systematic change of M/L with luminosity and
departure from homology were fully investigated. In order to preserve
the narrowness of the FP, the tilt can be explained fine tuning the
amount of dark matter to the galaxy luminosity, or fine tuning the
departures from homology to the galaxy size (e.g., \cite{ciotti96}).
Of the two options the former is the most widely accepted, and the
tilt ascribed to $M/L\propto L^{0.25}$ (e.g., \cite{bender93}).

Whatever the reason for the tilt is, both these possibilities face
paramount difficulties as soon as low-surface-brightness (LSB)
elliptical galaxies are considered, because these galaxies follows an
inverse trend $M/L\propto L^{-0.4}$ (\cite{dekel86};
\cite{peterson93}).  Thus, it look like any classical picture
conceived to explain the tilt of the FP defined by HSB, is deemed to
fail with LSB, and vice versa.  The goal of this work
is to investigate whether a consistent explanation for the tilt of the
FP and the opposite behavior of HSB and LSB galaxies can be ascribed
to a completely different physical phenomenon.

In 1983 a
modification of Newtonian dynamics known as MOND has been proposed as
an alternative to non-baryonic dark matter (Milgrom 1983). Milgrom's
proposal states that gravity (or inertia) does not follow the
prediction of Newtonian dynamics for acceleration smaller than
$a_0=1.2\times 10^{-8}$ cm~s$^{-2}$ \citep{begeman91}.  The true
acceleration of gravity $g$ being related to the Newtonian
acceleration $\gn$ by

\begin{equation}
\gn= g \mu(g/a_0).
\end{equation}

The interpolation function $\mu(g/a_0)$ admits the asymptotic
behavior $\mu(g/a_0)=1$ for $g>>a_0$, so to retrieve the Newtonian
expression in the strong field regime, and $\mu(g/a_0)=g/a_0$ for
$g<<a_0$. The constant $a_0$ is meant to be a new constant of physics.
 The transition from the Newtonian to the MOND regime 
occurs for values of acceleration that remain undetected within the solar
system where the strong field of the sun 
(of the order of $10^{-4}$ cm~s$^{-2}$ at the distance of Pluto)
dominates all dynamical processes.
As a consequence, we have no direct proof of the
validity of Newton's law in the low acceleration regimes that are
typical in galaxies.

Since the seminal papers on MOND \citep{milgrom83a, milgrom83b,
milgrom83c}, a few authors have worked on this subject showing how
this simple idea explains many properties of galaxies without the need
of non-baryonic dark matter
\citep[e.g.,][]{begeman91,milgrom94,sanders96,mcgaugh98}.  MOND is
also effective in describing the dynamics of galaxy groups and
clusters \citep{milgrom98,sanders99} and, with some approximations,
gravitational lensing \citep{qin95,mortlock01}. For a recent review
see \citet{sanders02}.

Clearly, MOND applies to  any stellar structure and therefore its
effects should be visible also in elliptical galaxies.
In this work, I shall study the FP  within the contest of MOND, to
see if it is possible to explain both the tilt of the FP 
and the opposite behavior of LSB and HSB galaxies 
without the needs to invoke non-baryonic dark matter.  

\section{MOND effects on elliptical galaxies}

\begin{figure}
\plotone{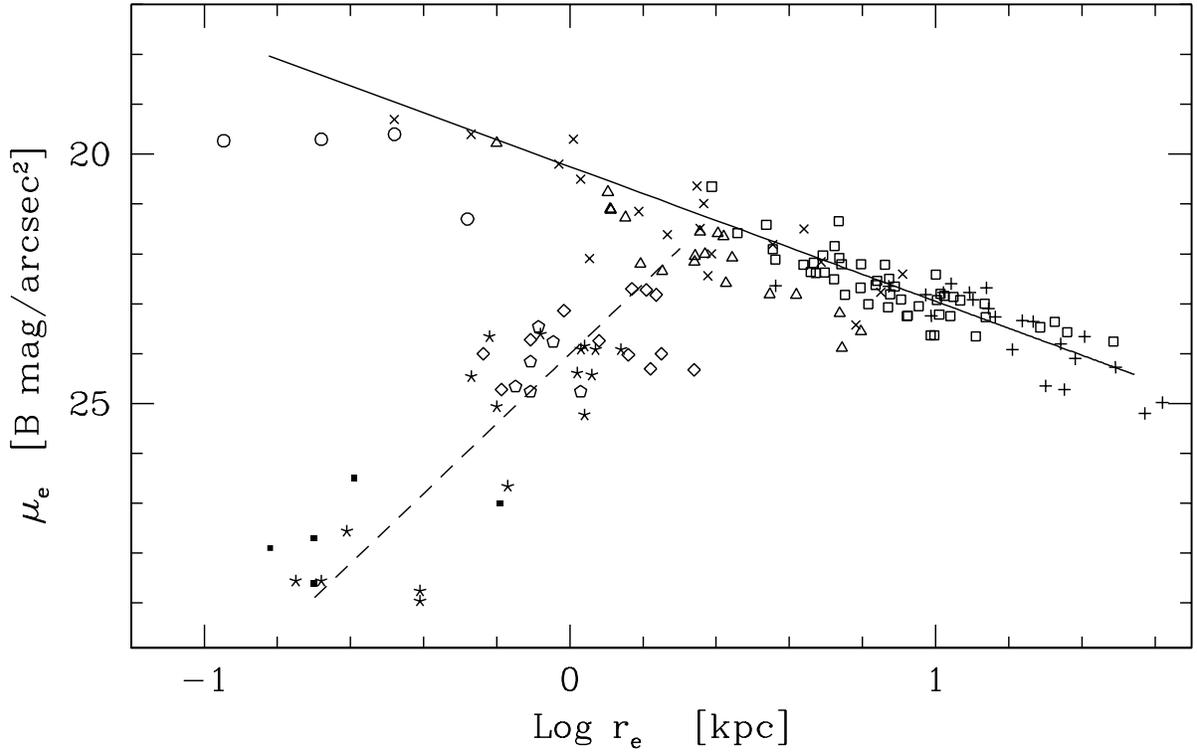}
\caption{Kormendy relation for various types of galaxies. Symbols are
as in Fig. ~\ref{FP}. 
HSB galaxies show a decreasing surface brightness as $r_e$ increases,
while  LSB show the opposite behavior.
The solid and dashed lines are an eye fit  to HSB and 
LSB galaxies, respectively, which have the sole purpose of better visualize 
the effects of MOND in the  following figures. 
\label{kormendy}
}
\end{figure}

If gravity deviates from Newtonian prediction in the low acceleration
regime typical of galaxies, then the logical implication is that the
weaker the field, the bigger the deviation. With this in mind, the
Kormendy relation (Fig. \ref{kormendy}) assumes a new meaning because
in elliptical galaxies $\Sigma$ (or
equivalently $\mu_e$), is proportional to their internal acceleration
of gravity
\footnote{ This derive from the assumption that there is no dark
matter.  In this case the acceleration at the effective radius is
$a=GxM/r_e^2$, where $x$ is the fraction of the total mass within
$r_e$.  After substituting $M=\tau L$ and $L=2\pi r_e^2 \Sigma$ we
have $a=2\pi G x \tau \Sigma \propto \Sigma$.}.  Thus the trends
followed by HSB and LSB galaxies in the $\mu_e -r_e$ plane 
can be seen as trends of their internal gravitational field. The
opposite dependence of $\mu_e$ on $r_e$ observed in HSB and LSB then
suggest these galaxies enter into MOND regime in different ways and,
as a result,  should follow different FP.
In order to use MOND to verify if this is the case,
the interpolation function $\mu$ must be specified.
Different but functionally equivalent expression for $\mu$ have been
used in the literature \citep[Milgrom 1983b, 1984;][]{sanders96}. 
Currently the most used function, also adopted here, is:

\begin{equation}
\mu(g/a_0) = \frac{ g/a_0}{\sqrt{1+(g/a_0)^2}},
\end{equation}

which works well in describing galactic rotation curves and
has the advantage of being simple.
Solving  for $g$  the true acceleration of gravity reads  

\begin{equation}
g=\gn \left[ \frac{1}{2}+\frac{1}{2}\sqrt{1+\left( \frac{2a_0}{\gn}\right)^2}\right]^{1/2}.
\end{equation}

This expression allows the calculation of the true gravitational 
acceleration provided the Newtonian one is known.

Before proceeding to compute $\gn$, it is worth noticing that
if $\tau$ is the true value of M/L of an object, and $\taun$ is 
the one derived using Newton's law to convert accelerations to masses, 
then $g/\gn \equiv \taun / \tau$. Indeed,
imagine an acceleration  $g$ is measured at distance $r$
from an object of luminosity $L$. In MOND we interpret this $g$ as due
to a mass $M=g r^2/G[]=\tau L$, where the term $[]$ is the factor $g/\gn$
defined in eq.3. On the other hand, according to Newton's law we
infer a mass $\Mn=g r^2/G = \taun L$. Then the ratio 
$\Mn/M=\taun/\tau = [] = g/\gn$. 
Thus, any difference between $g$ and
$\gn$ will  appear as a variation of $\tau$.

To investigate these effects on elliptical galaxies, I shall take as
representative of each galaxy the acceleration of gravity $g_N$ at $r_e$,
computed setting $\tau$ from
the stellar population typical of the galaxies 
and replacing the gravitational radius with $r_e$.
In absence of dark matter both assumptions are valid.
The mass within $r_e$ is set to be $x=0.41$ of the total mass, as
appropriate for galaxies following the $r^{1/4}$ de Vaucouleurs law
\citep{young76}.  Under these assumptions, the acceleration of
gravity at $r_e$ becomes:

\begin{equation}
\gn = \frac{G M x}{r_e^2} =
\frac{1.406\times 10^{-11} x \tau L}{r_e^2}  \;\;\;\;\left[cm~s^{-2}\right], 
\end{equation}

where the numerical constant is correct for $\tau$ and $L$ in solar units,
and $r_e$ in pc.
Combining eq. 3 and 4 one can compute $g/\gn$ for all galaxies. 

\section{The mass to light ratio}

Before proceeding with the calculation of $g_N$, I have to set the
value of $\tau$.
Since I am assuming Newton's law fails below $a_0$, masses derived
using any dynamical model based on this law can obviously not be used
here.

Stellar evolution theories, which potentially provide the best tools
for computing $\tau$, give results that critically depend on the adopted
initial mass function (IMF) and the low masses cutoff. These parameters
are rather uncertain and imply large differences on the derived
$\tau$. For instance, for a single coeval stellar population,
assuming a single-slope Salpeter IMF with slope 2.35,
$\tau$  changes by a factor
4.5 varying the low masses cutoff from 0.15 to 0.05 M$_{\odot}$. 
Note that this problem  is only mildly alleviated using a more realistic
multi-slope IMF \citep[e.g.,][]{scalo86} at low masses \citep{renzini93}.

An unbiased way to estimate $\tau$ relies on the analysis of the
stars in the solar neighborhood, where stars of all spectral type in all 
possible evolutionary phases can be studied in great detail. 
From \citet{binney98}, the
total mass of stars in  $10^4$ pc$^3$ is 356M$_{\odot}$ + 30M$_{\odot}$
of white dwarfs. For this sample it is found $\tau = 0.7$.

This sample of stars can be used to estimate $\tau$ for any
stellar population simply integrating the light over all the relevant
spectral types, while maintaining the total mass fixed.
This will result in an upper limit to the true $\tau$, because
more than one generation of stars contributed to it and therefore
the number of low mass stars is overrepresented.

In elliptical galaxies the stellar population is composed by stars in: 
sub-giant branch, red-giant branch, horizontal branch, and main
sequence up to spectral type K.  In $10^4$ pc$^3$ there
are 4 red giants of K type and 0.25 of M type populating the giant
branch, which have average luminosity of 57 and 120 solar
luminosities, respectively.  For each star in the giant branch there
are $\sim$ 21 stars in the sub-giant branch and 0.75 in the 
horizontal branch
\citep{renzini98}, which contribute for 5 and 52 $L_{\odot}$, 
respectively. Stars still in the main sequence add
another 18 $L_{\odot}$.  From these considerations it follows that
$\tau=386/420 \sim 0.9$.  
This value is consistent with that derived for globular clusters
\citep{mandushev91,renzini93}
 that have a similar old stellar population. 
These objects, however, are  slightly different form elliptical
galaxies because they don't contain gas and at least a fraction of the
less massive stars, which contributes to the mass but not to the
luminosity, have probably left the clusters long ago
\citep[e.g.][]{aguilar88, smith02}.  So, for globular cluster
$\tau=1$ should be fairly correct, while here I shall set $\tau=2$ for all
elliptical galaxies. A fairly conservative assumption that allows for
the presence of large amount of gas, particularly relevant in LSB
galaxies.

For spiral galaxies, setting $\tau=1.2$, that is about twice as big as 
that observed in the solar neighborhood, allows for
plenty of gas and the bulge old stellar population.  Finally, for
cluster of galaxies I assume a mix of spiral and ellipticals, with
$\tau \sim 1.6$.
I stress that the results reported in this work do not
significantly depend on the assumed value of $\tau$ in the range 1-2.  

\section{Comparing MOND with observations}

I compute here the quantity $g/g_N$ and investigate 
its relation with $\sigma$, $r_e$, and $L$ for a sample of
elliptical galaxies.

The representative data set for HSB galaxies is taken  from the
compilation of \citet{bender93}, to which I added data for 6 LSB and 11
dwarfs galaxies \citep{peterson93}, 6 compact ellipticals
\citep{marel02}, and 22 radio galaxies \citep{bettoni01}. 
Throughout the paper, H$_0=50$ km s$^{-1}$ Mpc$^{-1}$
is used.

\begin{figure}
\plotone{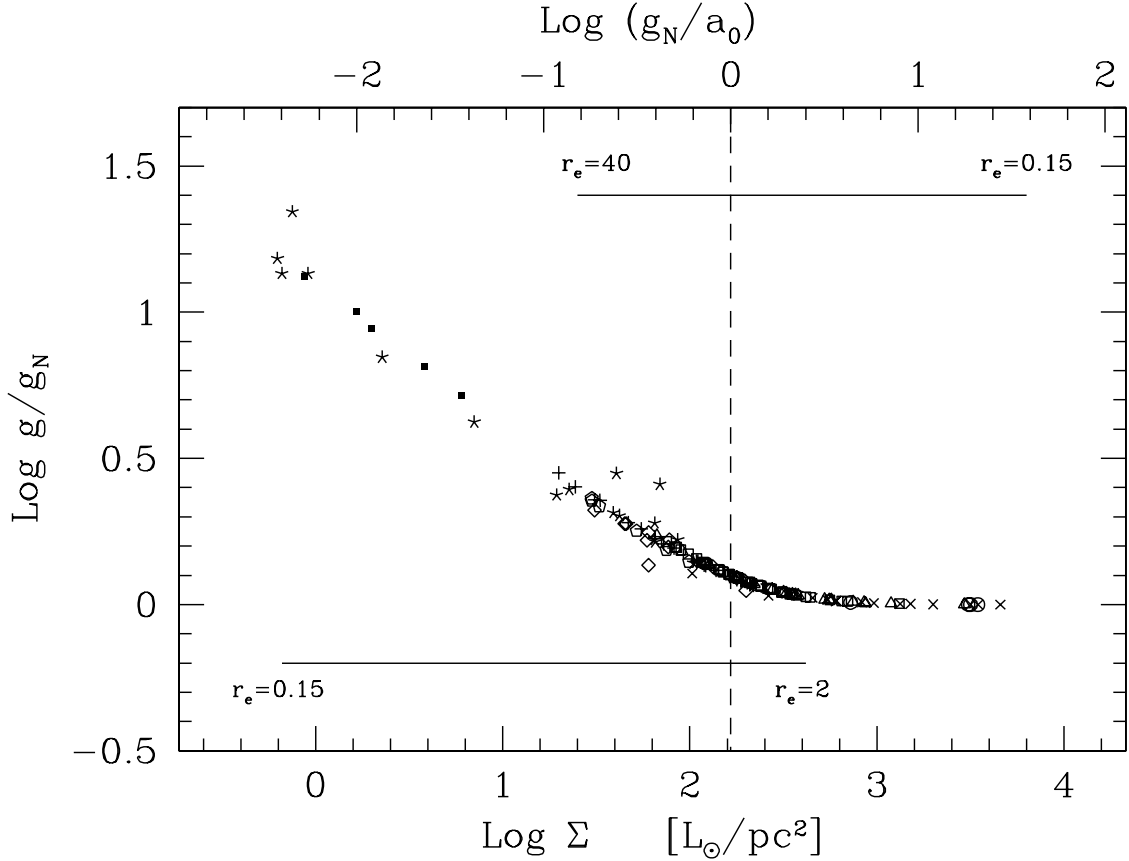}
\caption{Plot of $g/\gn \equiv \tau/\taun$ versus average surface
brightness $\Sigma$ for elliptical galaxies (symbols are as in
Fig. \ref{FP}).  The value of the Newtonian acceleration $\gn$ in
units of $a_0$ is also indicated in the upper scale, and the vertical
dashed line deviates the Newtonian from the MOND regime areas.  
The two horizontal lines give the
values of $r_e$ (in kpc) covered by HSB (upper line) and LSB (lower
line). Note that $r_e$ increases from left to right for LSB and vice
versa for HSB. This opposite dependence of $\Sigma$ on $r_e$ and the
corresponding variation of the mass-to-light ratio is causing the
different tilt of the FP defined by HSB and LSB galaxies.
}
\label{g_sigma}
\end{figure}

The acceleration $g_N$ at $r_e$ is found comparable to $a_0$ for HSB
galaxies, while being well below this value in LSB.  
The effect of MOND on objects characterized by different $\Sigma$
is shown in Fig. ~\ref{g_sigma}. It is seen that galaxies follow a
continuous sequence implying the mass discrepancy, or in other word
the dark matter content, is uniquely defined by $\Sigma$. I shall show
that this fact has important consequences for understanding why
objects as different as globular and galaxy clusters may appear to
have the same mass discrepancy.

\begin{figure}
\plotone{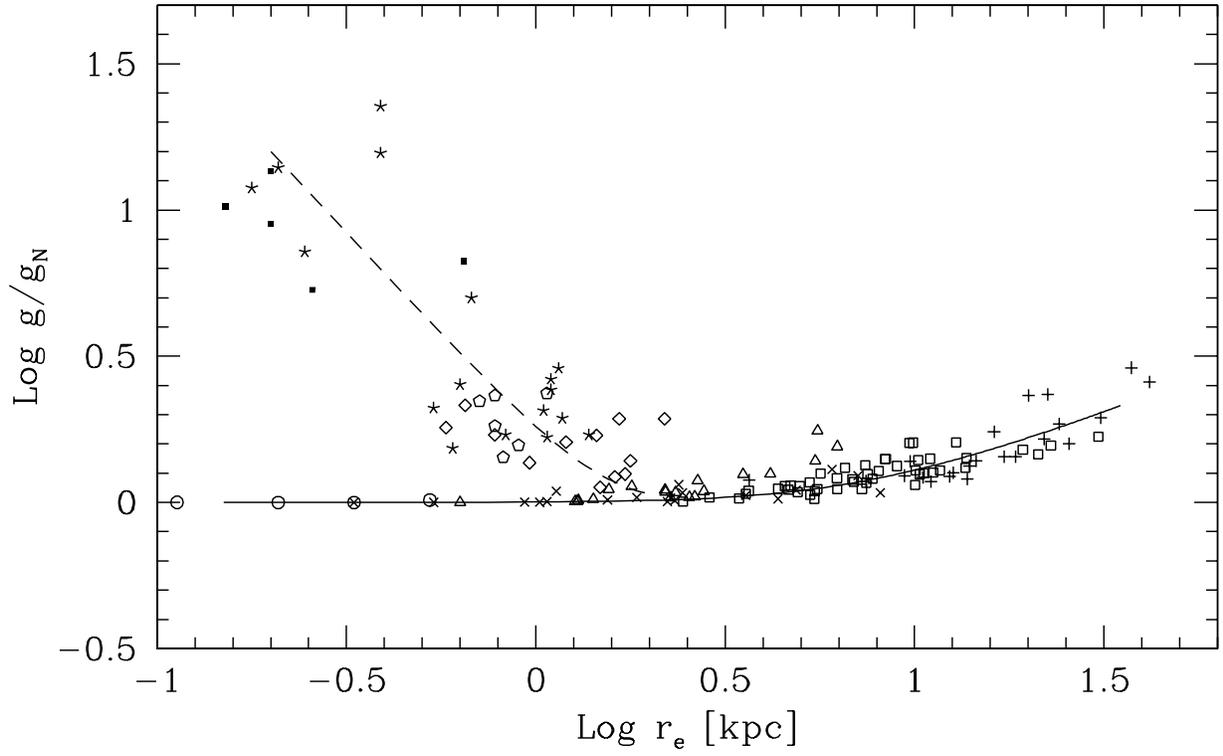}
\caption{Plot of $g/\gn \equiv \tau / \taun$ versus $r_e$.
Symbols are as in Fig. \ref{FP}. The opposite dependence 
of $\tau$ on $r_e$  shown by LSB and HSB galaxies is evident.
The dashed and solid lines are the 
same drawn in Fig. \ref{kormendy} and
are meant to visualize the loci followed by LSB and HSB galaxies.
}
\label{msul_re}
\end{figure}

\begin{figure}
\plotone{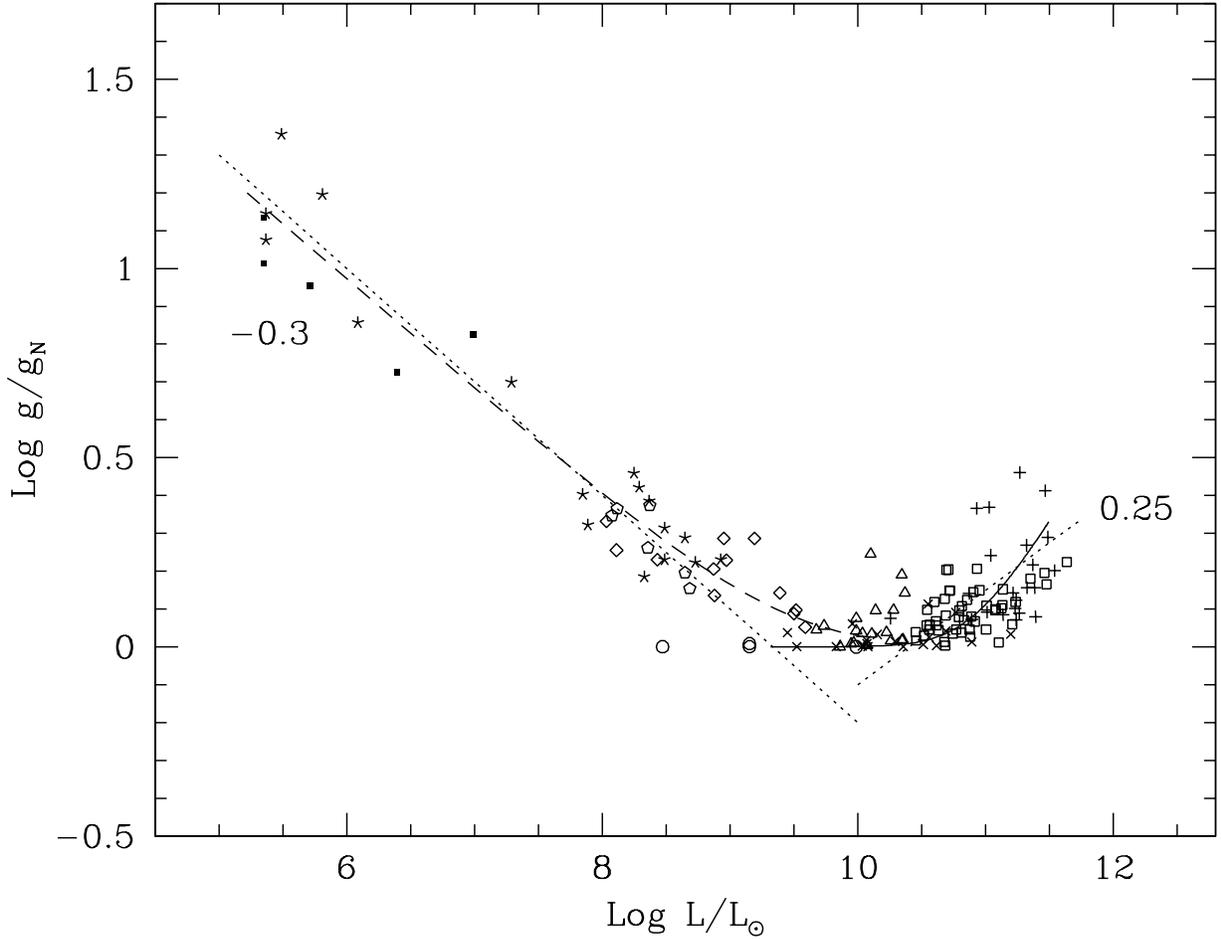}
\caption{
Plot of $g/\gn$ versus luminosity.
It is evident that $\tau / \tau_0$ decrease (increase) with luminosity
for LSB (HSB) galaxies. The dependence is
indicated by two dotted lines, which correspond to $\tau \propto L^{-0.3}$
in the case of LSB galaxies, and $\tau \propto L^{0.25}$ for HSB galaxies.
In absence of dark matter $k1 \propto L$, thus
this plot is the equivalent to Fig ~\ref{FP}. MOND therefore
predicts the right dependence of $\tau$ on $L$ for both HSB and LSB galaxies. 
The dashed and solid lines are the 
same drawn in Fig. \ref{kormendy} and
are meant to visualize the loci followed by LSB and HSB galaxies.}
\label{msul_lum}
\end{figure}

In Fig. ~\ref{g_sigma}, HSB and LSB galaxies are no longer
distinguishable because the opposite dependence of $\Sigma$ on $r_e$
(see Fig. \ref{kormendy}) disappears. The different behavior of HSB
and LSB become evident when plotting $g/\gn$ versus $r_e$ or $L$
(Fig. \ref{msul_re} and \ref{msul_lum}). It is therefore clear that
the effect of modified dynamics is to mimic an increase of $\tau$ as a
function of $L$ for HSB galaxies, while doing the opposite for LSB
galaxies. Moreover, Fig. \ref{msul_lum} shows that with good
approximation, and certainly within the spread of the data, the
dependence on luminosity is $\tau \propto L^{0.25}$ for HSB galaxies,
and $\tau \propto L^{-0.3}$ for LSB galaxies, consistent with the
values quoted in the literature \citep{bender93,peterson93}.

The striking different behavior of HSB and LSB galaxies can therefore
be easily explained by a unique model assuming a modification of
Newton's law. Under this hypothesis the tilt of the FP is not due to a
systematic increase/decrease of the dark matter content of galaxies,
but it is the effect of the variation of the strength of their
internal gravitational field as compared to $a_0$.

\begin{figure}
\plotone{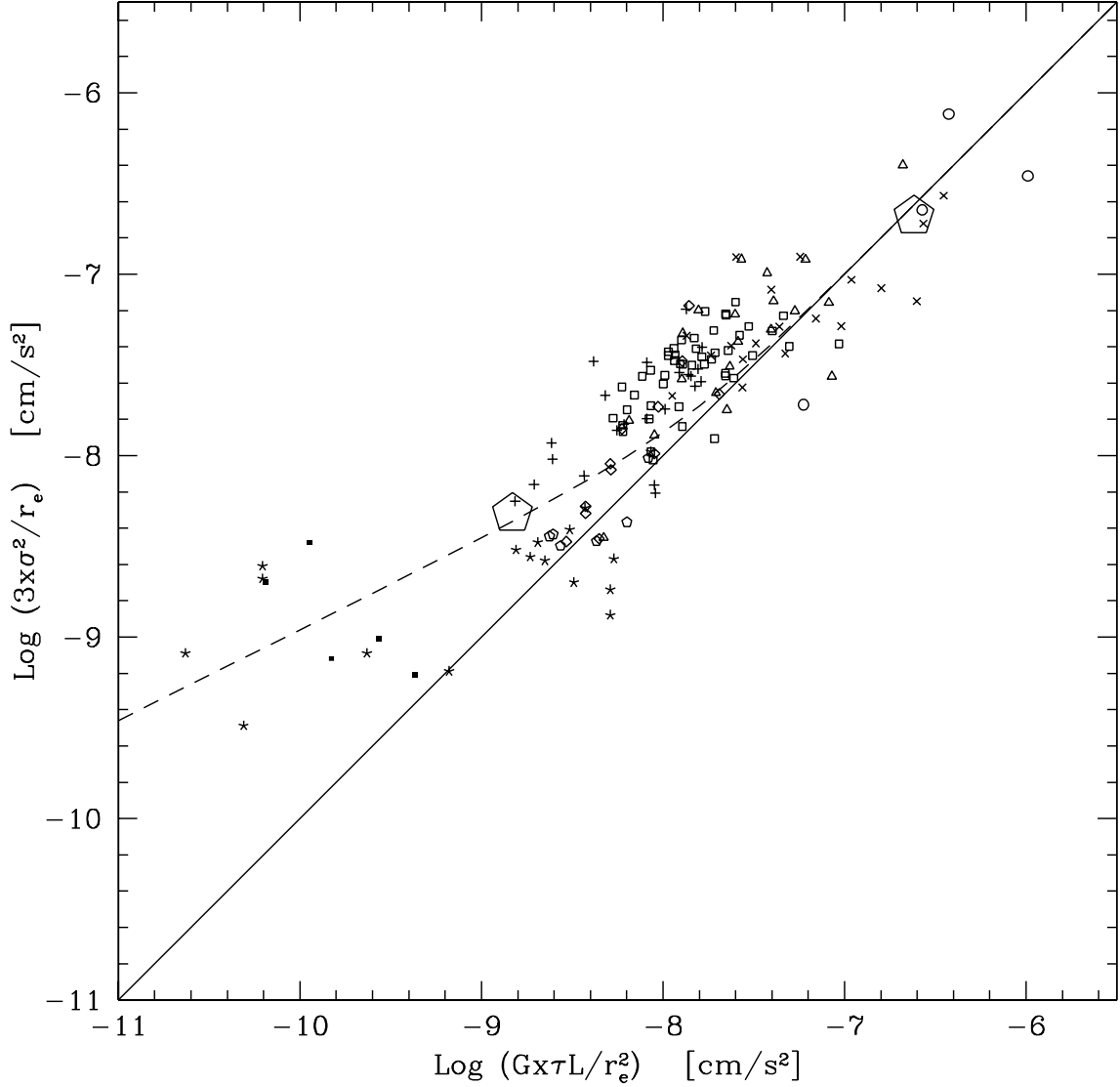}
\caption{Comparison of the gravitational acceleration at $r_e$ derived
using dynamical and luminous masses. The solid line gives the locus of
objects with no mass discrepancy, while the dashed line represents the
MOND prediction. Note the trend of increasing discrepancy as the
internal acceleration decrease. Symbols are as in Fig.~\ref{FP}}
except for the two big pentagons representing the globular cluster
Palomar 13 (left) and $\omega$ Centauri (right) discussed in the text.
\label{mondExp}
\end{figure}

Another way to investigate the effect of MOND on galaxies is shown in
Fig. \ref{mondExp}, where the
acceleration of gravity from eq. 4, is compared to the acceleration 
derived using the dynamical mass, $M_{dyn}=3G \sigma^2/r_e$. 
Though there is significant spread of the 
data, it is seen that dynamical masses imply no mass discrepancy for
acceleration stronger or comparable to $a_0$, while a deviation
is apparent when the acceleration goes below it, as prescribed by MOND.  
Note that the stronger discrepancy occurs for LSB galaxies, because 
they are in deep MOND regime. 
While observational data and MOND predictions agree reasonably well, 
to explain this behavior using Newtonian dynamics
requires fine tuning of the total amount of dark matter in galaxies 
of very different type and morphology.

\begin{figure}
\plotone{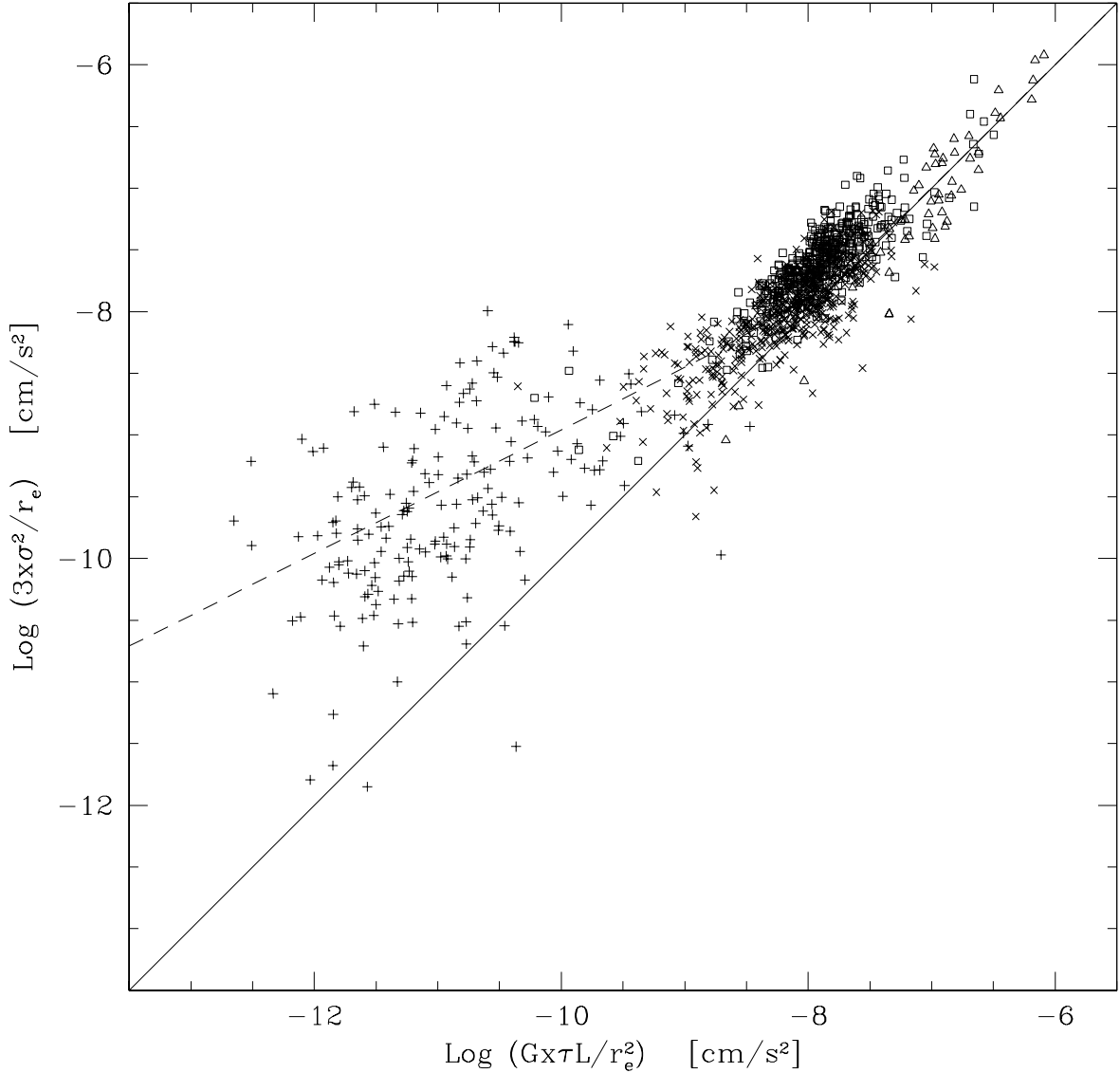}
\caption{\label{burstein1}
Same as figure \ref{mondExp} using data for 1050 objects from
\citet{burstein97}. Group and cluster of 
galaxies (pluses) are mostly on the left
very deeply in MOND, spiral galaxies (crosses) and elliptical galaxies
(squares) are at the center, and globular cluster (triangles), which
are mostly in Newtonian regime, are on the upper right.
Luminous masses where computed assuming $\tau=2$ for elliptical
galaxies, $\tau=1.0$ for globular clusters, $\tau=1.2$ for spiral
galaxies, and $\tau=1.6$ for group and cluster 
of galaxies (see text).}
\end{figure}

\begin{figure}
\plotone{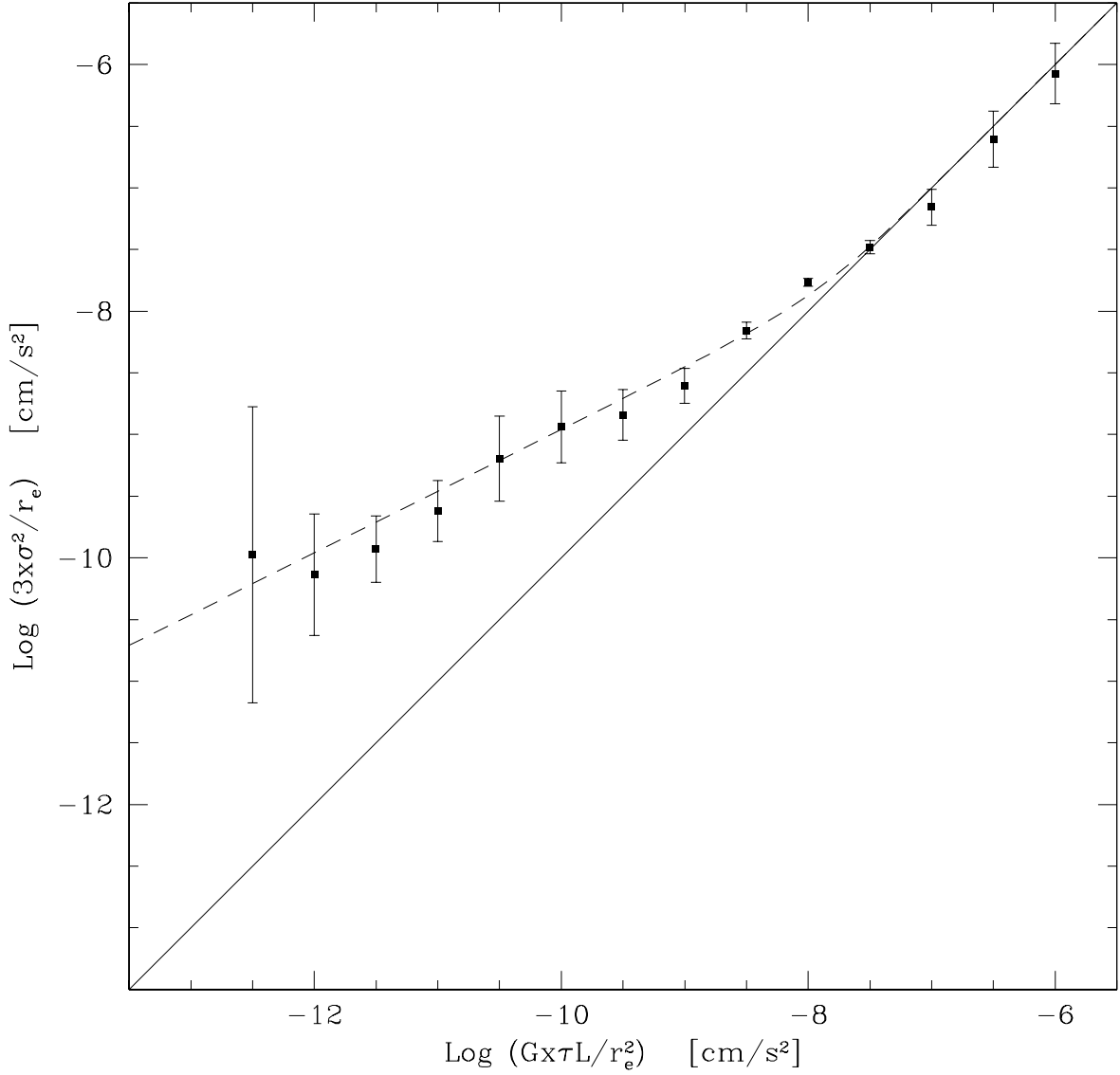}
\caption{\label{burstein2}
Same as Fig. \ref{burstein1} after re-binning the data in bins of 0.5 dex.
Error bars give $3\sigma$ confidence region of the average. 
In each bin, no distinction is made between objects of different type.
The agreement between observations and MOND is striking.
}
\end{figure}

These considerations are not limited to elliptical galaxies.  In fact
if MOND is the underlying law regulating gravity in the universe,
then all stellar systems should be described by MOND.  Though a full
discussion of this issue goes beyond the scope of the present work, it
is worth to add more data to Fig. \ref{mondExp} to see whether other stellar
systems different from elliptical galaxies follow the same trend.
In Fig. \ref{burstein1} I plot the data from the data for 1050 objects
(51 globular clusters, 300 elliptical galaxies, 513 spiral
galaxies, 170 galaxy groups and 16 galaxy clusters) from the
compilation of  \citet{burstein97}, are plotted.  
Data cover 7 order of magnitudes in internal acceleration, going from
globular clusters which are believed to be dark matter free, to the
largest cluster which have $\tau \sim 100$.  The agreement between
MOND and the observed dependence of the mass discrepancy is
clear and becomes striking after re-binning the data
(Fig. \ref{burstein2}), indicating that either the dark-matter
content in stellar structure is amazingly well linked to the strength
of their internal gravitational field, 
or that Newton's law fails below $a_0$.

I stress that under the latter explanation, it is 
clear that objects like globular clusters are baryons dominated not
just for a coincidence but because their internal acceleration of
gravity is above $a_0$. And objects as different as a galaxy cluster
and a tiny LSB dwarf galaxy have similarly large $\tau$ not for a
pure coincidence but because their internal acceleration of gravity is
similar.

Finally, I would like to comment on recent result for two globular
clusters. Observations of the stellar velocity dispersion of Palomar
13 \citep{cote02} have been interpreted as evidence for large amount
of dark matter in this cluster ($\tau \sim 40$).  Similarly, my
collaborators and I have shown that at large radii the velocity
dispersion profile of $\omega$ Centauri remains flat for accelerations
below few times $a_0$ \citep{scarpa02}, as observed in galaxies. This
result was interpreted as evidence of a failure of Newton's law in the
low acceleration limit, while the classical scenario would imply large
amount of dark matter surrounding the cluster and therefore large mass
discrepancy.  I show in Fig. \ref{mondExp} that in both cases the
clusters uniforms to MOND prediction further supporting the proposed
scenario.

\section{Conclusion}
The most important conclusion of this work is that
assuming MOND the tilt of the FP can be
explained by the different strength, as compared to $a_0$, of the
internal gravitational field in galaxies of different type and size.
This suffice to show that using Newton's law of gravity $\tau$ is
expected to vary systematically with $r_e$ or, equivalently, with $L$,
in a way fully consistent with observations. The opposite behavior of
HSB and LSB galaxies, which poses paramount difficulties for Newtonian
dynamics, also find a natural explanation in this scenario.  
The observed systematic deviation extends naturally to other
stellar structures over 7 order of magnitude in acceleration, 
strongly suggesting that a law of physics, rather than dark matter,
regulates the behavior of stellar structures in the
Universe. It is worth noticing that all this is achieved in MOND
without a single free parameter.\\

\noindent
{\bf Acknowledgments}\\
It is a pleasure to thanks Renato Falomo and Gianni Marconi,
for helpful discussion that strongly improved the quality of this work.

\clearpage

\clearpage

\end{document}